\documentclass{article}
\usepackage{graphicx}
\usepackage{subfig}
\usepackage{caption}
\usepackage{amsmath}
\usepackage{amssymb}
\usepackage{amsthm}
\usepackage[margin=1.0in]{geometry}
\newcommand{\OPTT}{T^*}
\newcommand{\OPTAR}{\tau^*}
\newcommand{\MAXAR}{\tau^0}
\newcommand{\Greedy}{\tau^G}
\DeclareMathOperator\size{area}
\DeclareMathOperator\weightActR{v}
\DeclareMathOperator\weightEvent{w}
\newcommand{\uweight}{weight}
\newcommand{\uweights}{weights}
\newcommand{\eweight}{volume}
\newcommand{\eweights}{volumes}
\newcommand{\dweight}{total volume} % diagram weight

\usepackage{hyperref}
\usepackage{todonotes}
\usepackage[ruled]{algorithm2e}
\usepackage[normalem]{ulem}

\newcommand{\activitydiag}{T}

\newtheorem{theorem}{Theorem}
\newtheorem{corollary}{Corollary}
\newtheoremstyle{break}
  {\topsep}{\topsep}%
  {\itshape}{}%
  {\bfseries}{}%
  {\newline}{}%
\theoremstyle{break}
\theoremstyle{break}
\newtheorem*{problem*}{\textsc{TimeWindowLabeling}}

\newcommand{\annikachange}[1]{#1}
\newcommand{\hermanchange}[1]{{#1}}

\title{Analysis of a Greedy Heuristic for \\ the Labeling of a Map with a Time-Window Interface}
\author{Annika Bonerath, Anne Driemel, Jan-Henrik Haunert,\\ Herman Haverkort, Elmar Langetepe, Benjamin Niedermann}

\begin{document}
\maketitle
%\herman{To do: inform Elmar and Benjamin about this paper that has their names on it}
%\herman{To do: I was still a bit worried that there might be confusion between ``weight'' (= value per unit of area in the configuration space) and ``effective weight'' or ``weighted activity region size'' (= weight times area in the configuration space) and ``total weight'' (total effective weight in the whole diagram), but I am not sure about better wording yet. Maybe ``unit weight'', ``region weight'', and ``total diagram weight''? I have now defined macros \texttt{uweight}, \texttt{eweight}, and \texttt{dweight} and used them throughout the document, so that we can change his easily. It needs another proofread to check if this is now all correct and consistent.}
%\annika{check: analyze - analyse, ...}
\section*{Abstract}
In this paper, we analyze the approximation quality of a greedy heuristic for automatic map labeling. As input, we have a set of events, each associated with a label at a fixed position, a timestamp, and a \uweight. Let a \emph{time-window labeling} be a selection of these labels such that all corresponding timestamps lie in a queried time window and no two labels overlap. %\herman{I edited what follows a bit, to make it clear that we study solutions in which we the labeling of each time window is fixed in advance and cannot depend on the query history, and to clarify what it is we are approximating}
A solution to the time-window labeling problem consists of a data structure that encodes a time-window labeling for each possible time window; when a user specifies a time window of interest using a slider interface, we query the data structure for the corresponding labeling. 

We define the quality of a time-window labeling solution as the sum of the \uweights\ of the labels in each time-window labeling, integrated over all time windows. We aim at maximizing the quality under the condition that a label may never disappear when the user shrinks the time window. 
In this paper, we analyze how well a greedy heuristic approximates the maximum quality that can be realized under this condition.

On the one hand, we present an instance with square labels of equal size and equal \uweight\ for which the greedy heuristic fails to find a solution of at least 1/4 of the quality of an optimal solution. On the other hand, we prove that the greedy heuristic does guarantee a solution with at least 1/8 of the quality of an optimal solution. In the case of disk-shaped labels of equal size and equal \uweight, the greedy heuristic gives a solution with at least 1/10 of the quality of an optimal solution. 
%\herman{added:}
If the labels are squares or disks of equal size and the maximum \uweight\ divided by the minimum \uweight\ is at most $b$, then the greedy heuristic has approximation ratio $\Theta(\log b)$.

\section{Introduction}
\begin{figure}[t]
    \centering
    \subfloat[data]{
    \includegraphics[page=1]{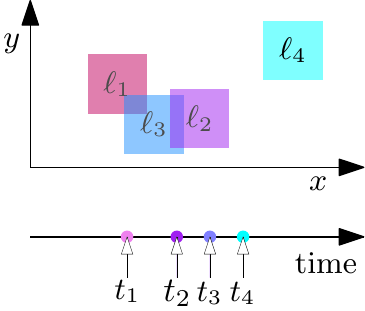}
    \label{fig:data}
    }\hfill
    \subfloat[user interface]{
    \includegraphics[page=2]{figures/arxiv-problemdef.pdf}
    \label{fig:userinterface}
    }\hfill
    \subfloat[basic interactions]{
    \includegraphics[page=3]{figures/arxiv-problemdef.pdf}
    \label{fig:basicinteractions}
    }
    \caption{Problem setting for events $e_1,e_2,e_3,$ and $e_4$ with timestamps $t_1,t_2,t_3,$ and $t_4$ and labels $\ell_1,\ell_2,\ell_3,$ and $\ell_4$ and a time-window query $Q = [t',t'']$ that contains the events $e_2,e_3,$ and $e_4$. The two overlap-free labels $l_3$ and $l_4$ form a time-window labeling of $Q$ (marked with a black stroke in (b)). The user can change the time-window query with four basic interaction as illustrated in (c). }
    \label{fig:problemdef}
\end{figure}

\begin{figure}[tb]
    \centering
    \subfloat[flickering]{
    \includegraphics[page=4]{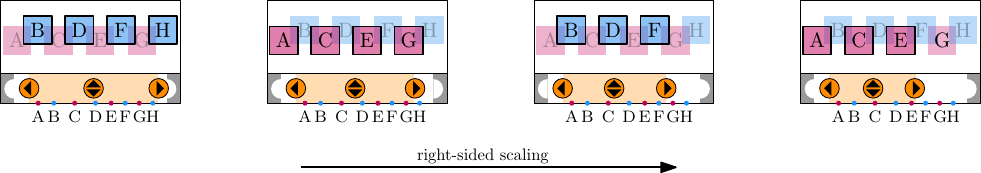}
        \label{fig:flickering}
    }
    
    \subfloat[without flickering]{ 
    \includegraphics[page=5]{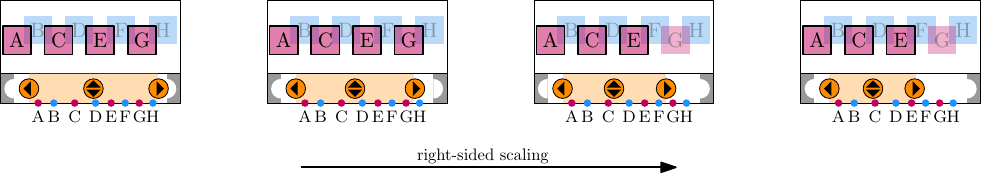}
        \label{fig:noflickering}
    }
    \caption{Sequence of map frames for one basic interaction. The labels highlighted with a black stroke are displayed, while the other labels and the timestamps are shown only for illustration.}
    \label{fig:stability}
\end{figure}
For the visualization of spatial data, labeling is a standard technique. Labels are placed in the map over the data points and each label contains information about the corresponding data point, e.g., a name or an icon. Typically, placing all labels leads to unwanted overlaps. Classical map labeling solves this by computing a largest overlap-free selection of labels~\cite{Yoeli1972}. Labeling is also often used for interactive maps. Here the user interactively changes the visualization and therewith the labeling must be updated. Recent research concerns stability and consistency conditions for labeling during a change in the visualization~\cite{Zhang2020, Bahrdt2017, Peng2020, Meijers2020}. 

\annikachange{
Bonerath et al.~\cite{bonerath2022algorithms} look at consistent labeling of maps with a time-slider interface. They introduce a consistency model and describe a data structure that guarantees such consistency criteria. Furthermore, they provide algorithms for the computation of the data structure. In this document, we provide the theoretical analysis of the greedy heuristic presented by Bonerath et al.~\cite{bonerath2022algorithms}. In the following, we wrap up their application scenario, model, data structure and algorithm. 
}

% our data + interface
\paragraph{Application Scenario}
In this work, we consider event data as input where each event consists of a label, a timestamp and a \hermanchange{positive} \uweight, where the \uweight\ of a label reflects its importance; see \autoref{fig:data}. Our application scenario consists of a user interface where the user can interactively choose a time window  and then an overlap-free  selection of labels with corresponding timestamps in the queried time window is visualized; see \autoref{fig:userinterface}. We call such a selection of labels a \emph{time-window labeling}. The users can choose the time window with a dynamic query interface as introduced by Williamson and Shneiderman~\cite{doi:10.1057/palgrave.ivs.9500097}. They can perform four \emph{basic interactions}; 
%\herman{Rephrased a bit to make it easier to parse} 
see \autoref{fig:basicinteractions}: (1) \emph{panning}: a continuous translation of the time window; (2) \emph{uniform scaling}: a continuous change of both boundaries of the time window in opposite directions, such that the center of the time window remains the same; (3+4) \emph{right- and left-sided scaling}: a continuous change of the time window's right or left boundary, respectively.

% computation and problem setting
\paragraph{Model}
As introduced by Bonerath et al.~\cite{bonerath2022algorithms}, we look at a two-step approach: 
%\herman{Rephrased a bit to clarify, see comment in abstract} 
first, we compute a data structure that encodes a time-window labeling for each possible time window, and then, as the user specifies time windows of interest, we query the data structure for the corresponding labelings. 
%\herman{Inserted here, since otherwise the reference to Been et al. cannot be understood:}
An alternative view of the data structure is the following: a query with a time window $[t',t'']$ can be regarded as a point $(t',t'')$ in a two-dimensional \emph{configuration space}: the first coordinate of the point specifies the starting time of the time window, the second coordinate specifies the end time. The data structure encodes, for each label $\ell$, its \annikachange{\emph{activity region}} $\tau$: the set of points $(t',t'')$ in configuration space such that $\ell$ is included in the labeling for the time window $[t',t'']$. A query with a time window $[t',t'']$ consists of finding all labels whose activity regions include the point $(t',t'')$.

As for classic map labeling, we aim at transferring as much information as possible for each time-window query. In particular, we want to maximize the sum of the s\ of displayed labels integrated over all time-window queries. This is inspired by active range maximization as introduced by Been et al.~\cite{Been2006}. 
%We call this property \textsc{MaxInfo}. 
A naive approach on computing the labelings for this interface would be to apply the classic strategy (find a largest overlap-free selection of labels) independently for each  possible time window. This might lead to unwanted flickering effects: a single label may appear and disappear repeatedly even within a single basic interaction; see \autoref{fig:flickering}. One effect of flickering is that the user cannot isolate a single event by systematically shrinking the time window, as the corresponding label may appear and disappear repeatedly  without any recognizable systematic. We require that if an event is displayed for a time window $Q=[t',t'']$ then it is also displayed for all the time windows that are contained in $Q$ and contain the timestamp of the event. \autoref{fig:noflickering} shows a solution that satisfies this requirement. We call this property \textsc{Containment}.

\paragraph{Related Work}
\emph{Map Labeling} is a widely investigated field. For the static case, a common goal is to maximize the number of the displayed labels (or their total \uweight) while avoiding overlapping labels~\cite{Agarwal1998, Yoeli1972, haunert2017beyond}. For non-interactive animated maps, additional stability constraints are added~\cite{bnns-tmlan-16,Gemsa2020,Bobak2020}. For interactive maps, Been et al.~\cite{Been2006} introduced the concept of active ranges 
%\herman{We call them \emph{activitiy regions}. I guess we should either make the terminology consistent, or make it explicit that Been's active ranges and our acitivity regions are basically the same concept} 
for labels, considering zooming, panning and rotations of the map \annikachange{which is basically the same concept as our activity regions. They consider the active range to be an interval over, e.g., zoom levels and prove that for such a scenario the maximization of all active ranges is NP-hard. }
%\herman{under the \textsc{Containment} constraint?} 

The data structure that is discussed in this paper, can be classified as a  \emph{time-windowed data structure}. This concept from the field of computational geometry, and it summarizes data structures that aim at efficiently answering time-window queries. Our approach is a time-windowed data structure that uses labeling as the underlying visualization technique. Nevertheless, in general, time-windowed data structures do not consider any consistency criteria during interaction. Current research on time-windowed data structures focuses on relational event graphs~\cite{Bannister2013, Chanchary2019, Chanchary2019b}, basic problems from computational geometry~\cite{Bannister2014, Bokal2015, Chan2015, Chan2016, Chanchary2018}, and also on event visualization based on $\alpha$-shapes \cite{Bonerath2019} and density maps \cite{Bonerath2020}. 

\emph{Approximation algorithms} are efficient algorithms that provide solutions for problems with a guarantee for the quality of the solution with respect to the optimal solution. They are often developed for NP-hard problems. The approximation ratio is a measure for the quality of an algorithm. Let $\textrm{OPT}(I)$ denote the optimal solution of a maximization problem for instance $I$ and ${\cal A}(I)$ the solution computed by algorithm $\cal A$. Then, $\cal A$ has \emph{approximation ratio} $k$ if $\textrm{OPT}(I)/{\cal A}(I) \leq k$ for all instances $I$ of the problem. For minimization problems, one can define concepts analogously. 
%\herman{I'd say that for an audience of computer scientists, references to text books on approximation algorithms are not needed}
%For more details, we want to refer to~\cite{vazirani2001approximation,williamson2011design,hochba1997approximation,johnson1974approximation}. 
In this paper, we discuss the approximation ratio of our greedy heuristic. 

\annikachange{
\paragraph{Our results}
In this paper, we discuss the approximation ratio of the greedy heuristic presented by Bonerath et al.~\cite{bonerath2022algorithms}. In \autoref{sec:formalization}, we give a formalization of the problem and provide a detailed description of the approximation ratio of the greedy heuristic in \autoref{thm:greedyapxratio}. Then, the results from \autoref{sec:lowerboundsapprox} and \autoref{sec:upperboundapprox} together prove \autoref{thm:greedyapxratio}.
In detail, in \autoref{sec:lowerboundsapprox}, we discuss the lower bound of the approximation ratio of the greedy heuristic. In \autoref{sec:instance1}, we give an exemplary instance for which the greedy heuristic has an approximation ratio above 4. In \autoref{sec:instance2}, we give a family of instances for which the approximation ratio is above $n/2$, where $n$ is the number of the input events. We provide a deeper analysis of this family of instances in \autoref{sec:refconstrlowerbound}, leading to a more accurate lower bound. In \autoref{sec:upperboundapprox}, we discuss the upper bound of the approximation ratio. }

%% -------------------------------------------------------------
%%
\section{Problem formalization and algorithm}\label{sec:formalization}

\paragraph{Formalization}
%\herman{Added: definition of a label}
A \emph{label} is a set of points in the plane, for example, a square, a rectangle or a disk with a specific location.
Let $\{\ell_1,\hdots, \ell_n\}$ be a set of labels, $\{t_1,\hdots, t_n\}$ be a set of timestamps, and  $\{\weightEvent_1,\hdots,\weightEvent_n\}$ a set of \hermanchange{positive} \uweights. We call the triplet $e_i = (l_i,t_i,\weightEvent_i)$ for $1\leq i \leq n$ an event. The input data for the algorithm and the data structure is a set of events $E$.
%Let $E=\{e_1,\dots,e_n\}$ be a set of events with corresponding labels $\ell_1,\hdots, \ell_n$, timestamps $t_1,\hdots, t_n$, and ratings $\weightEvent_1,\hdots,\weightEvent_n$. 
%\herman{Why specify weights as a function while specifying labels and timestamps with simple indices? Moreover, it is a bit confusing that we use $w$ (in some way or another) for both relative weights (here) and absolute weights (weight times area, in the description of the algorithm below). Proposal: replace $w(e_i)$ by $\weightEvent_i$ (for relative weight). Alternative proposal: replace $w_i$ by $v_i$ (for value) or $p_i$ (for potential) and replace $w(e_i)$ by $w_i$.}
We say that two events and their labels are in \emph{conflict} if their labels overlap, 
%\herman{Made the definition of overlap precise:}
that is, their interiors have a non-empty intersection. The dynamic query interface introduces two additional input parameters, the minimal and maximal time slider positions $t_{\text{min}}$ and $t_{\text{max}}$. We call a range $Q=[t',t''] \subseteq [t_{\text{min}}, t_{\text{max}}]$ a \emph{time-window query}. Be aware that depending on the context we interpret $Q$ either as an interval $[t',t'']$ or a point $(t',t'')$ in the plane (configuration space). Due to $t' \leq t''$ it holds that $(t',t'')$ always lies in the triangle $(t_{\textrm{min}},t_{\textrm{min}})$, $(t_{\textrm{max}},t_{\textrm{max}})$, $(t_{\textrm{min}},t_{\textrm{max}})$. We say that if a label $\ell_i$ is contained in a time-window labeling of $Q$, then $e_i$ (and also $\ell_i$) is \emph{active} for $Q$. Let $\tau_i$ be the set of time-window queries for which $e_i$ is active. We call $\tau_i$ \emph{activity region} of $e_i$. Analogously to time-window queries, we understand $\tau_i$ as a subset of $\mathbb R^2$. We call $\activitydiag = \{\tau_1,\hdots, \tau_n\}$ \emph{activity diagram} if for each pair of events that are in conflict the corresponding activity regions do not overlap. Hence, querying an activity diagram  $\activitydiag$ with $Q$ corresponds to reporting all events where their activity region in $\activitydiag$ contains $Q$. We receive a time-window labeling. \annikachange{We call $\weightActR(\tau_i) = \weightEvent_i\cdot \mathrm{area}(\tau_i)$ the \emph{\eweight\ of an activity region} $\tau_i$ where $\mathrm{area}(\tau_i)$ is the area of $\tau_i$. We also call $\weightActR(\tau_i)$  the \emph{\eweight\ of event} $e_i$. Further, we introduce $\weightActR(\activitydiag)=\sum_{i=1}^n \weightActR(\tau_i)$ as the \emph{\dweight\ of the activity diagram} $\activitydiag$.} An \emph{optimal} activity diagram is one that has maximum \dweight\ among all activity diagrams that satisfy \textsc{Containment}.
%Maximizing $w(\activitydiag)$ leads to property \textsc{MaxInfo}. 
%\herman{The definition of \textsc{MaxInfo} confuses me. It does not seem to be independent of \textsc{Containment}, because it seems I should understand \textsc{MaxInfo} as: having maximum weight subject to \textsc{Containment}? If \textsc{MaxInfo} would be defined independent of \textsc{Containment}, then realizing \textsc{Containment} will usually mean that you do not get \textsc{MaxInfo}. Is there really a need to define \textsc{MaxInfo}? Can we not simply say: an \emph{optimal} activity diagram is one that has maximum weight among all activity diagrams that satisfy \textsc{Containment}?}

It is easy to see 
%\herman{Well, in the other paper it is a Lemma that needs a proof of half a column, so let's refer to that explicitly. Where do we actually refer to the other paper for the first time? Should happen in Section 1 or even before.}
that the activity region $\tau$ in an optimal activity diagram  of an event $e$ at timestamp~$t$ must be a rectangle with lower right corner~$(t,t)$ (Lemma 1 in Bonerath et al.~\cite{bonerath2022algorithms}). \hermanchange{Furthermore, the definition of \textsc{Containment} implies that any activity diagram that consists of such rectangles satisfies \textsc{Containment}. Thus we arrive at the following problem formulation:}

\begin{figure}[tb]
    \centering
    \includegraphics[page=6]{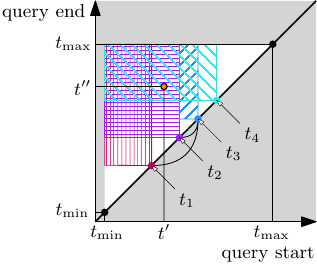}
    \caption{Valid activity diagram for the data presented in \autoref{fig:problemdef}. Pairs of conflicting events are marked with an arc, i.e., the pair $e_1$ and $e_3$ and the pair $e_2$ and $e_3$. The time-window query $Q=[t',t'']$ intersects the activity regions of $e_2$ and $e_4$.}
    \label{fig:activity-diagram}
\end{figure}

\begin{problem*}\label{problem:timewindowedlabeling} \vspace{0.5em}
 \begin{tabular}{ll}
     \textbf{Given:} & A set~$E=\{e_1,\dots,e_n\}$ of spatio-temporal events with labels; a weighting function $w\colon E\to \mathbb R^+$; \\& the bounds $t_\mathrm{min}$ and $t_\mathrm{max}$ of the activity diagram.\\
     \textbf{Find:} & An activity diagram $\activitydiag=\{\tau_1,\dots,\tau_n\}$ of activity regions for $E$ that maximizes  $\sum^n_{i=1} \weightEvent_i\cdot \mathrm{area}(\tau_i)$ \\& for $e_i\in E$, where $\mathrm{area}(\tau_i)$ is the area of $\tau_i$ in the activity diagram and where $\tau_i$ is a \\ & rectangle with lower right corner $(t_i,t_i)$.  \\
\end{tabular}
\end{problem*}
%\hermanchange{\sout{We say that an optimal solution for \textsc{TimeWindowLabeling} is an \emph{optimal} activity diagram;}}\todo{Herman: removed, because optimal activity region has been defined above already, we cannot redefine it here} 
\hermanchange{For an example of an optimal activity diagram,} see \autoref{fig:activity-diagram}.
%\todo{Herman edited this figure to handle comment by Elmar: confusion between ``query end'' and $t_{\max}$. Please check Figs 1,2,4 if nothing broke in the file. -- annika: looks good} 
%\herman{I believe the diagram in the figure is not optimal if the events all have the same relative weight. I guess it is optimal if $e_3$ is heavy enough, so I guess we should specify the weights.}

\begin{figure}[tb]
    \centering
    \includegraphics[page=7]{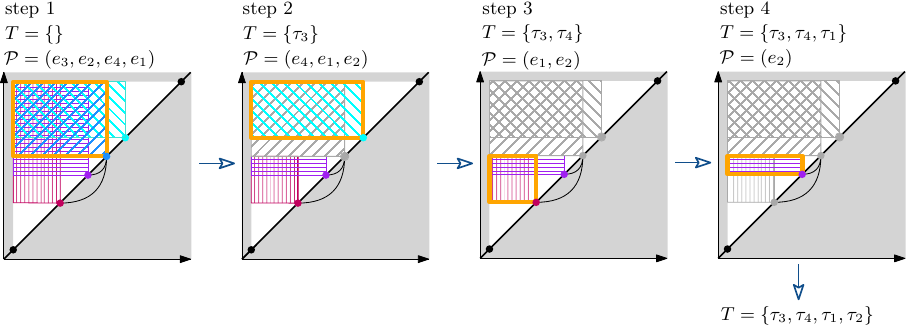}
    \caption{Greedy heuristic for the data presented in  \autoref{fig:problemdef}. For each step, the activity region that is chosen is marked with an orange stroke.}
    \label{fig:greedy}
\end{figure}

\paragraph{Algorithm}
Next, we present our greedy heuristic for computing a valid activity diagram. For illustration see \autoref{fig:greedy} and \autoref{alg:greedy}. The greedy heuristic successively selects activity regions that yield the largest gain. While doing so, it maintains for each event that has not yet been placed in the activity diagram its maximal potential activity region. Each time a new event is selected and placed in the diagram, all remaining activity regions that are in conflict with this event are trimmed and their potential contribution is updated accordingly.

\begin{algorithm}[tb]
\caption{Greedy Heuristic for \textsc{TimeWindowLabeling}}\label{alg:greedy}
\KwData{events $e_1,\hdots,e_n$}
%\herman{Should also give the labels, timestamps and weights as data}
\KwResult{activity diagram $\activitydiag$}
initialize the solution set $\activitydiag = \{\}$\;
%\herman{Rewrote a bit, because a priority queue is not usually maintained by completely sorting its contents, as that would be a waste of computation time}
initialize an empty priority queue ${\mathcal Q}$ to hold events organized by decreasing \eweight\;
\For {$i=1\hdots n$}{
%\herman{Corrected definition of the initial activity region written as a Cartesian product:}
    initialize activity region $\tau_i \gets [t_{\textrm{min}},t_i]\times [t_i,t_{\textrm{max}}]$\;
    initialize the \eweight\ $\weightActR(\tau_i) \gets \weightEvent_i\cdot \mathrm{area}(\tau_i)$\;
    insert $e_i$ in ${\mathcal Q}$ with \eweight\ $\weightActR(\tau_i)$
}
\While{$\activitydiag$ is not empty}{
  extract element $e_i$ of maximum \eweight\ from ${\mathcal Q}$ and add $e_i$ to $T$\;
%\herman{Corrected the next lines: just cutting $\tau_i$ out of $\tau_j$ may have the result that $\tau_j$ is no longer a rectangle, we need to cut a bit more aggressively. I believe cutting out the \emph{original} $\tau_i$ (given that $\tau_j$ intersects the \emph{current} $\tau_i$), does the trick: to be sure, please compare to the code of the implementation}
  \ForEach {$e_j \in {\mathcal Q}$ such that $\ell_j$ overlaps $\ell_i$ and $\tau_j$ overlaps $\tau_i$}{
    $\tau_j \gets \tau_j \setminus [t_{\textrm{min}},t_i]\times [t_i,t_{\textrm{max}}]$\;
    update \eweight\ $\weightActR(\tau_j) \gets \weightEvent_j\cdot \mathrm{area}(\tau_j)$ and update ${\mathcal Q}$ accordingly
  }
}
\end{algorithm}

More in detail, we initialize for each event $e_i\in E$ its largest possible activity region $\tau_i$, i.e., the region that is spanned by $(t_{\mathrm{min}},t_{\mathrm{max}})$ and $(t_i,t_i)$ and further, its \emph{\eweight} $\weightActR(\tau_i)$. We initialize a priority queue~$\mathcal Q$ of events in descending order by their \eweights\ and the empty solution set~$\activitydiag$; see step 1 in \autoref{fig:greedy}. Then, we remove the first event $e_i$ from $\mathcal Q$ (with largest \eweight) and add $\tau_i$ to the solution set~$\activitydiag$. For each remaining event~$e_j$ in $\mathcal Q$ that is in conflict with $e_i$ we trim $\tau_j$ to the largest possible activity region~$\tau'_j\subseteq \tau_j$ that does not intersect $\tau_i$. Finally, we update the \eweight\ of $e_j$ to $\weightEvent_j\cdot\mathrm{area}(\tau'_j)$, possibly changing the position of $e_j$ in the sorting of~$\mathcal Q$. We iterate until $\mathcal Q$ is empty. Finally, we return the valid activity diagram $\activitydiag$.

\annikachange{
\paragraph{Approximation Ratio}
Let $E$ be a set of events. Let $E_i \subseteq E$ be the set of events that are in conflict with $e_i \in E$. Then, let $a_i$ be the maximum size of a subset of $E_i$ where no two events are in conflict. Let $a$ be the maximum over all $a_i$ with $1\leq i \leq n$.  Let $b \in \mathbb R$ such that for any two events $e_i, e_j \in E$, we have $1/b \leq \weightEvent_i/\weightEvent_j \leq b$. We call $a$ the \emph{degree of interference} of $E$ and $b$ the \emph{degree of unbalance} of $E$. Using the degree of interference and the degree of unbalance, we can describe the approximation ratio of the greedy heuristic as follows. 
}
\begin{theorem}\label{thm:greedyapxratio}
Let $E$ be a set of events with degree of interference $a$ and degree of unbalance $b$. The approximation ratio of the greedy heuristic is $\Theta(\min(a \log b, n))$. If $b=1$, that is, all labels have equal \uweight, then the approximation ratio is at most $2a$.
\end{theorem}

%% ----------------------------------------------------------------------
\section{Lower bounds on the approximation ratio of the greedy heuristic}\label{sec:lowerboundsapprox}
\subsection{An instance with approximation ratio above 4}\label{sec:instance1}
\begin{figure}
    \subfloat[input labels]{\leavevmode\kern-8mm
    \includegraphics[page=1,scale=0.69]{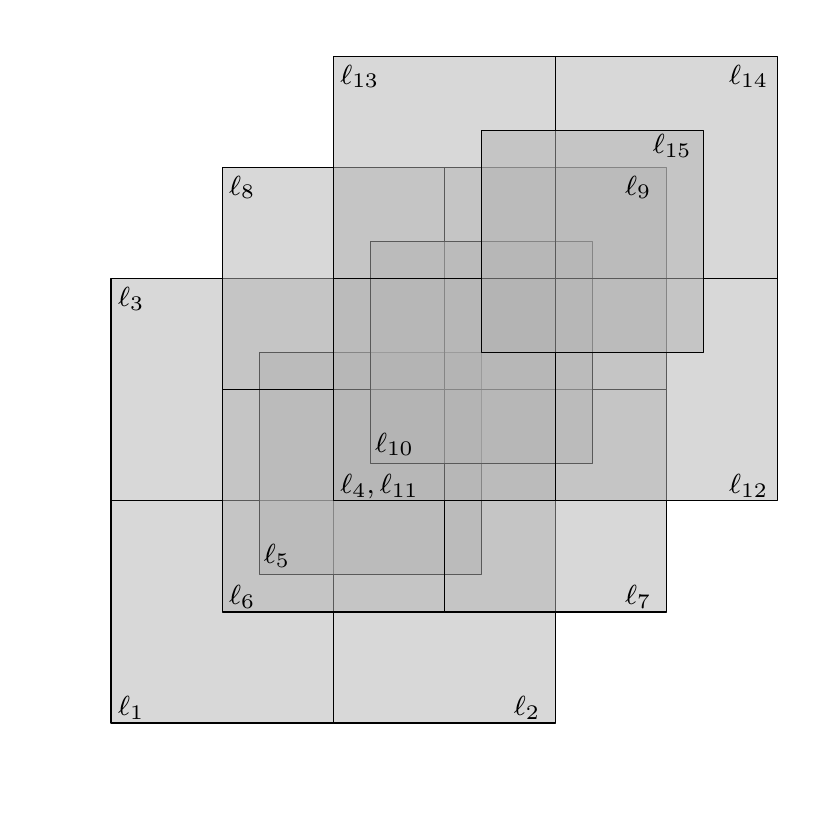}%
    }%
    \subfloat[activity diagram]{% 
    \includegraphics[page=2,scale=0.69]{figures/arxiv-badgreedy.pdf}%
    }%
    \subfloat[activity diagram from greedy]{% 
    \includegraphics[page=3,scale=0.69]{figures/arxiv-badgreedy.pdf}%
    }
    \caption{(a) A bad input for the greedy heuristic. (b) A solution \hermanchange{(not true to scale)} with total active region size $900 + 26\varepsilon - 7\varepsilon^2$. Each part of the activity diagram is labelled with the numbers of the labels whose activity region covers that part. (c) The greedy solution \hermanchange{(not true to scale)} with total active region size only $207 + 107\varepsilon - 13\varepsilon^2$.}
\label{fig:greedylowerbound}
\end{figure}

In this section, we provide an instance where the approximation ratio of the greedy algorithm is above 4.
\paragraph{Instance}
The instance consists of a set of $n = 15$ events of equal \uweight\ with square-shaped labels of size $6 \times 6$. \autoref{tab:instance-above-4} specifies their  centre points and timestamps. For an illustration see  \autoref{fig:greedylowerbound}a. Note that two labels overlap each other if and only if their centre points differ by less than 6 in both coordinates.
We consider the minimal query starting time to be $t_{\textrm{min}}= 0$ and the maximal query ending time to be $t_{\textrm{max}}=24$. 

\paragraph{Optimal Solution} A valid activity diagram is shown in \autoref{fig:greedylowerbound}b, with total active region size $900 + 26\varepsilon - 7\varepsilon^2$, a lower bound for the optimal solution. 

\begin{table}[tb]
\begin{centering}
\begin{tabular}{lcc|lcc|lcc|lcc|lcc}
& centre & $t$ &
& centre & $t$ &
& centre & $t$ &
& centre & $t$ &
& centre & $t$ \\\hline
$\ell_1$ & (0,0) & 8 &
$\ell_2$ & (6,0) & 8 &
$\ell_3$ & (0,6) & 8 &
$\ell_4$ & (6,6) & 8 &
$\ell_5$ & (4,4) & $8+2\varepsilon$ \\
$\ell_6$ & (3,3) & 16 &
$\ell_7$ & (9,3) & 16 &
$\ell_8$ & (3,9) & 16 &
$\ell_9$ & (9,9) & 16 &
$\ell_{10}$ & (7,7) & $16+\varepsilon$ \\
$\ell_{11}$ & (6,6) & 21 & 
$\ell_{12}$ & (12,6) & 21 &
$\ell_{13}$ & (6,12) & 21 &
$\ell_{14}$ & (12,12) & 21 &
$\ell_{15}$ & (10,10) & $21-\varepsilon$ \\
\end{tabular}
\end{centering}
\caption{Instance where greedy heuristic has approximation ratio above 4. The parameter $\varepsilon$ is an arbitrarily small positive number between 0 and 1/34.}
\label{tab:instance-above-4}
\end{table}

\paragraph{Greedy Solution}
The greedy heuristic, however, would first place $\ell_5$ with activity region $[0,8+2\varepsilon] \times [8+2\varepsilon,24]$ of size $128 + 16\varepsilon - 4\varepsilon^2$. Note that $\ell_5$ intersects all labels $\ell_1,...,\ell_{11}$. Thus, in the activity diagram:\begin{itemize}
\item $\ell_1,...,\ell_4$ are now confined to the rectangle $[0,8] \times [8,8+2\varepsilon]$ of size $16\varepsilon$;
\item $\ell_6,...,\ell_9$ are now confined to the rectangle $[8+2\varepsilon,16] \times [16,24]$ of size $64-16\varepsilon$;
\item $\ell_{10}$ is now confined to the rectangle $[8+2\varepsilon,16+\varepsilon] \times [16+\varepsilon,24]$ of size $64-16\varepsilon + \varepsilon^2$;
\item $\ell_{11}$ is now confined to the rectangle
$[8+2\varepsilon,21] \times [21,24]$ of size $39-6\varepsilon$;
\item $\ell_{12},...,\ell_{14}$ may still get active regions $[0,21] \times [21,24]$ of size 63;
\item $\ell_{15}$ may still get an active region $[0,21-\varepsilon] \times [21-\varepsilon,24]$ of size $63+18\varepsilon-\varepsilon^2$.
\end{itemize}
Therefore, the greedy heuristic would now select $\ell_{10}$, which intersects all other labels except $\ell_1,...,\ell_3$. Thus, the activity regions of the remaining labels are now restricted as follows:
\begin{itemize}
\item $\ell_1,...,\ell_4$ are still confined to the rectangle $[0,8] \times [8,8+2\varepsilon]$ of size $16\varepsilon$;
\item $\ell_6,...,\ell_9$ are now confined to the rectangle $[8+2\varepsilon,16] \times [16,16+\epsilon]$ of size $8\varepsilon-2\varepsilon^2$;
\item $\ell_{11},...,\ell_{14}$ are now confined to the rectangle
$[16+\varepsilon,21] \times [21,24]$ of size $15-3\varepsilon$;
\item $\ell_{15}$ is now confined to the rectangle $[16+\varepsilon,21-\varepsilon] \times [21-\varepsilon,24]$ of size $15-\varepsilon-2\varepsilon^2$.
\end{itemize}
Thus, the next label selected by the greedy heuristic is $\ell_{15}$, after which $\ell_1,...,\ell_4$ get activity regions $[0,8] \times [8,8+2\varepsilon]$ of size $16\varepsilon$; the labels $\ell_6,...,\ell_9$ get activity regions $[8+2\varepsilon,16] \times [16,16+\epsilon]$ of size $8\varepsilon - 2\varepsilon^2$, and $\ell_{11},...,\ell_{14}$ get activity regions $[21-\varepsilon,21] \times [21,24]$ of size $3\varepsilon$.

Thus, the greedy heuristic achieves a total activity region size of $
(128 + 16\varepsilon - 4\varepsilon^2) + 
(64-16\varepsilon + \varepsilon^2) + 
(15-\varepsilon-2\varepsilon^2) +
4 \cdot 16\varepsilon +
4 \cdot (8\varepsilon - 2\varepsilon^2) +
4 \cdot 3\varepsilon =
207 + 107\varepsilon - 13\varepsilon^2$.

\paragraph{Approximation Ratio} Thus, the optimal solution beats the greedy heuristic by a factor of at least:\[
\frac{900 + 26\varepsilon - 7\varepsilon^2}{207 + 107\varepsilon - 13\varepsilon^2}.
\]
For small $\varepsilon$ this factor approaches $900/207 = 100/23$. This is not tight; slightly shifting the timestamps of the instance could make the ratio slightly worse still.

%% ----------------------------------------------------------------------
\subsection{A family of instances with approximation ratio at least $n/2$}\label{sec:instance2}
\begin{figure}
    \centering
    \includegraphics[page=1]{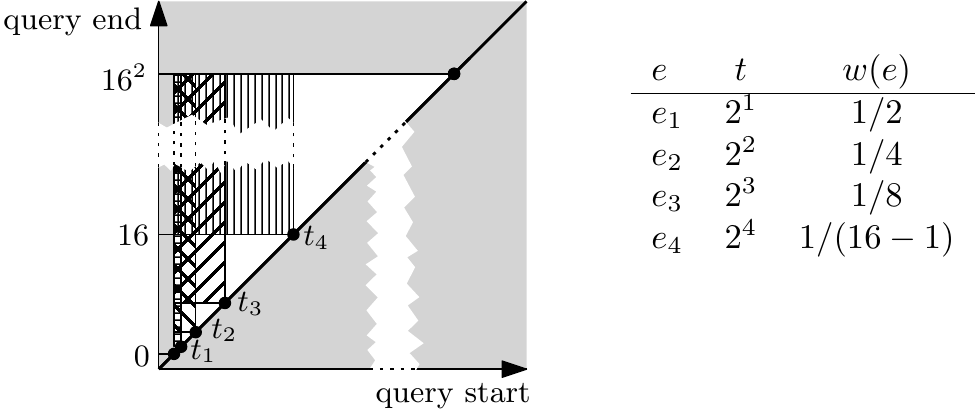}
    \caption{Instance illustrating Section~\ref{sec:instance2} for $b= 16$.}
\label{fig:instance2}
\end{figure}
\paragraph{Instances}
With events of different \uweights\ we can even construct input instances that cause the greedy heuristic's performance to become arbitrarily bad. 
%\herman{In this section we should maybe look at the terminology a bit more carefully, for consistency with the previous sections, e.g. I guess that, instead of "maximum weighted activity region size", we could, depending on the context, maybe just say "quality", or "maximum effective weight" etc.}
Choose an interval $[1,b]$ from which to pick the \uweights, such that $b$ is an integral power of two, larger than 1. \annikachange{Let $n$ be $\log_2 b$. Let $t_\textrm{min}=0$ and $t_\textrm{min}=b^2$.} We create $n$ events $e_1,...,e_n$ in the time window $[0, b^2]$, where $e_n$ has \uweight\ $\frac{1}{b-1}$; the events $e_j$, for $j \in \{1,...,n-1\}$, have \uweight\ $2^{-j}$; each event $e_j$, for $j \in \{1,...,n\}$, has timestamp $2^j$, and all labels have the same location; see \autoref{fig:instance2} for $b=16$.%\todo{Herman changed the figure to confirm that $16^2$ is really a lot bigger than 16.} 
\paragraph{Optimal Solution}
Note that $e_n$ has maximum \eweight\ $\hermanchange{\frac{1}{b-1} 2^n (b^2 - 2^n) = \frac{1}{b-1} b (b^2 - b) =\ } b^2$, whereas each other event $e_j$ has maximum \eweight\ $\hermanchange{2^{-j} 2^j (b^2 - 2^j) =\ } b^2- 2^j$. The optimal solution would contain at least the right half of each event's maximum possible region (and for $e_1$, also the left half); the \dweight\ will be roughly $\frac12 nb^2$. More precisely, the \dweight\ of this solution would indeed be:\[
\frac{b^2}{2} + \sum_{j=2}^{n-1}\frac{b^2 - 2^j}{2} + \left(b^2 - 2\right) = 
\frac{(n+1)b^2}{2} - 2^{n-1} = \frac{(n+1)b^2 - b}{2} = \frac{n b^2 + b(b-1)}{2} > \frac{n b^2 + n b}{2}.\]
In the last step, we used $b-1 \geq \log_2 b = n$.

\paragraph{Greedy Solution} The greedy heuristic, however, would first give $e_{n}$ its maximum possible region. This reduces the maximum height of the activity region of each other event $e_j$ from $b^2 - 2^j$ to $2^n - 2^j = b - 2^j$; thus its maximum \eweight\ is reduced to:\[
2^{-j} (b - 2^j) = 2^{-j} b - 1 = 2^{n-j} - 1 < 2^{n-j},
\]
and the maximum total \eweight\ of all events is reduced to less than:\[
b^2 + \sum_{j=1}^{n-1} 2^{n-j} < 
b^2 + 2^n = b^2 + b
\]

\paragraph{Approximation Ratio} Thus, the greedy heuristic's solution is worse than the optimal solution by a factor of at least:\[
\frac{(n b^2 + n b)/2}{b^2 + b} = \frac{n}{2}.
\]

%\herman{Added:}
Note that the factor $n/2$ is reached under the condition $n = \log_2 b$, or conversely, $b = 2^n$. In other words, the construction requires events whose \uweight\ differences are exponential in $n$. Where this is not realistic, the lower bound might better be expressed in terms of $b$, as we do in the next subsection.

\subsection{A refined construction of the lower bound}\label{sec:refconstrlowerbound}
%\herman{(for lack of a better title) I would put this here, to have the lower bounds all together}
We can extend the construction given above to labels that do not all have the same location. 
\hermanchange{Fix numbers $a \geq 1$ and $b \geq 1$, where $b$ is $2^m$ for some integer $m$; we will construct a set $E$ of $n = (a+1) m$ events with degree of interference $a$ and degree of unbalance $b$.}
%\todo{Herman: replaced $n$ by $m$ it what follows}
%Let $E_i \subseteq E$ be the set of events that is in conflict with $e_i \in E$. Then, let $a_i$ be the maximum size of a subset of $E_i$ where no two events are in conflict. Let $a$ be the maximum over all $a_i$ with $1\leq i \leq n$. In other words, $a$ is the maximum size of a set of mutually disjoint labels that all intersect a common label that is not in the set. Note that if all labels are unit squares, then $a \leq 4$ (one label on each corner of the common intersected label); if all labels are unit disks, then $a \leq 5$. Let $b \in \mathbb R$ such that for any two events $e_i, e_j \in E$, we have $1/b \leq \weightActR(e_i)/\weightActR(e_j) \leq b$. 
%\herman{Where did this come from? The other manuscript mentions it, right?}
%For a lower bound on the approximation ratio, we use a similar strategy as for the instances presented in Section~\ref{sec:instance2}. 
Concretely, let $E$ consist of $(a+1) \times m$ events $e_{i,j}$, for $i \in \{0,...,a\}$ and $j \in \{1,...,m\}$, with timestamps in the time window $[0, b^2]$. For all $i \in \{0,...,a\}$, event $e_{i,m}$ has \uweight\ $\frac{1}{b-1}$; the events $e_{i,j}$, for $j \in \{1,...,m-1\}$, have \uweight\ $2^{-j}$; all events $e_{i,j}$, for $j \in \{1,...,m\}$, have timestamp $2^j$. For each $i$, the labels $\ell_{i,j}$ have the same locations, such that $\ell_{0,j}$ intersects all other labels $\ell_{i,j}$, but these labels do not intersect each other. Thus, two different events $e_{g,h}$ and $e_{i,j}$ are in conflict if and only if $g = 0$, $i = 0$, or $g = i$. Note that the maximum \eweight\ for any event $e_{i,j}$ with $j \in \{1,...,m-1\}$ is $2^{-j} 2^j (b^2 - 2^j) = b^2 - 2^j$; the maximum \eweight\ for any event $e_{i,m}$ is $\frac{1}{b-1} 2^m (b^2 - 2^m) = \frac{1}{b-1} b (b^2 - b) = b^2$.

\paragraph{Optimal Solution}
\hermanchange{There is a solution that places}, at each point of the diagram, the $a$ events of highest \uweight\ that are in range and are not in conflict with each other. The \dweight\ is thus:\begin{eqnarray*}
&& a 2^{-1} 2^1 (b^2 - 2^1) + a \left(\sum_{j=2}^{m-1} 2^{-j} (2^j - 2^{j-1}) (b^2 - 2^j)\right) + a \frac{1}{b-1} (2^m - 2^{m-1}) (b^2 - 2^m) \\
&& =(a/2) ((m+1) b^2  - b)  \\
&& =\Omega(a b^2 \log b).
\end{eqnarray*}

\paragraph{Greedy Solution}
The greedy heuristic however, could start with giving event $e_{0,m}$ its maximum region, thus eliminating the events $e_{i,m}$ for $i \in \{1,...,a\}$ completely, and reducing the maximum size of the other labels' activity regions by a factor at least $b$. Moreover, in the following steps, the greedy heuristic could always pick an event $e_{0,j}$, thus eliminating $e_{i,j}$ for $i \in \{1,...,a\}$. In the end, the greedy solution will have \dweight\ at most:
\begin{eqnarray*}
&& 2^{-1} 2^1 (2^m - 2^1) + \left(\sum_{j=2}^{m-1} 2^{-j} (2^j - 2^{j-1}) (2^m - 2^j)\right) + b^2  \\
&& = (1/2) (2b^2 + mb - b)  \\
&& = O(b^2).
\end{eqnarray*}

\paragraph{Approximation Ratio}
Thus, the approximation ratio of the greedy heuristic is at least $\Omega(a \log b)$ in the worst case\hermanchange{, which proves} the lower bound stated in \autoref{thm:greedyapxratio}.
\hermanchange{Note that with a given number of events $n$, the lower bound construction can only be realized as long as $(a+1)\log b \leq n$, since the construction requires this many labels. If $(a+1)\log b > n$, we can only do the construction for a smaller degree of interference $a'$ and a smaller degree of interference $b'$ such that $(a'+1)\log b' = \Theta(n)$, and the approximation ratio of the greedy heuristic is $\Omega(a' \log b') = \Omega(n)$. Thus, the lower bound on the worst-case approximation ratio is $\Omega(a \log b)$ or $\Omega(n)$, whatever is lower.}
In the next section we will see that this lower bound is tight up to constant factors. 
%% ----------------------------------------------------------------------
\section{Upper bound on the approximation ratio of the greedy heuristic}\label{sec:upperboundapprox}
In this section, we derive an upper bound on the approximation ratio of the greedy heuristic and hence\annikachange{, together with the results of \autoref{sec:lowerboundsapprox}, we prove  \autoref{thm:greedyapxratio}.}
%We recap that the approximation ratio is the worst-case ratio between on the one hand, the maximum total weight of the activity regions of a valid labeling, and on the other hand, the total weight of the activity regions selected by the greedy heuristic. 
%\herman{We can formulate this more concisely if we give $a$ and $b$ names and define them above. Maybe the \emph{degree of interference} and the \emph{degree of unbalance}?}
%\begin{theorem}\label{thm:greedyapxratio}
%Let $E=\{e_1,\dots,e_n\}$ be a set of events with a weighting function $w\colon E\to \mathbb R^+$. Let $E_i \subseteq E$ be the set of events that is in conflict with $e_i \in E$. Then, let $a_i$ be the maximum size of a subset of $E_i$ where no two events are in conflict. Let $a$ be the maximum over all $a_i$ with $1\leq i \leq n$. Let $b \in \mathbb R$ such that for any two events $e_i, e_j \in E$, we have $1/b \leq \weightActR(e_i)/\weightActR(e_j) \leq b$. The approximation ratio of the greedy heuristic is $\Theta(\min(a \log b, n))$. If $b=1$, that is, all labels have equal weight, then the approximation ratio is at most $2a$.
%\end{theorem}
%In the following, we prove \autoref{thm:greedyapxratio}. 
%\subsection*{Upper Bound}
Without loss of generality, let the time scale run from 0 to~1.
%\herman{The optimal solution may not be unique, so we have to be careful that in the definition of $\OPTAR$, we do not mix different optimal solutions. Added:}
Let $\OPTT(E)$ be an arbitrary optimal solution for $E$. Let $e_i \in E$ and $\OPTAR_i$ be the activity region of $e_i$ in $\OPTT(E)$, let $\Greedy_i$ be the activity region of $e_i$ in the greedy solution, and let $\MAXAR_i$ be the maximum possible activity region of $e_i$, that is, the rectangle $[0,t_i] \times [t_i,1]$. 
%For an activity region $\tau_i$ of $e_i$, let $\weightActR(\tau_i)$ be its weighted size $\weightEvent_i \cdot \size(\tau_i)$. 
Our goal is now to determine an approximation ratio, that is, to determine a factor $k$ (ideally as low as possible) such that the following holds for any set of events $E$:
\begin{equation*}
    \frac{\sum_{e_i \in E}\weightActR(\OPTAR_i)}{\sum_{e_i \in E}\weightActR(\Greedy_i)} \leq k.
\end{equation*}
%\herman{Notation of weights is consistent ($w$ versus $\weightActR$)}

\paragraph{Charging:}
In order to prove the upper bound, we need to introduce the concept of charging. Let $e_i$ be an event in $E$. \hermanchange{From now on, we define each event to be in conflict with itself.} Let $e_j$ be the first-placed event in the greedy solution that is in conflict with $e_i$ 
and whose active region $\Greedy_j$ intersects $\MAXAR_i$, \annikachange{that is, among all events of $E$ that are in conflict with $e_i$, the event $e_j$ is the first \hermanchange{to be extracted from the priority queue by} the greedy heuristic.}
\hermanchange{Such an event $e_j$ always exists; it might be $e_i$ itself.} 
%\herman{Check: is it clear what first-placed means? (first in the order of extraction from the priority queue)}
We say $e_i$ \emph{charges} $\weightActR(\OPTAR_i)$ to $e_j$. 
%\hermanchange{\sout{In other words, for $e_i $, the event $e_j$ is the first event selected by the greedy heuristic such that $e_j$ is in conflict with $e_i$.}} 

\hermanchange{By this charging, we model the following circumstances:} 
Before $e_j$ is selected, the event $e_i$ could still get $\MAXAR_i$ as its activity region. However, \hermanchange{before (or when) the greedy heuristic selects $e_i$, it selects $e_j$}, so we know we must have $\weightActR(\MAXAR_i) \leq \weightActR(\Greedy_j)$. 
%\herman{Added:}
After selecting $e_j$, the event $e_i$ cannot get an(other) activity region of size $\MAXAR_i$ anymore. We ``blame'' $e_j$ for that by charging $\weightActR(\OPTAR_i)$ to $e_j$. \hermanchange{Note that the amount charged is only $\weightActR(\OPTAR_i)$, not $\weightActR(\MAXAR_i)$.}
% OLD 
%The proof is based on a charging scheme: any label $\ell_i$ charges $\weightActR(\OPTAR_i)$ to the label $\ell_j$ that is the first-placed label in the greedy solution that intersects $\ell_i$ and whose active region $\Greedy_j$ intersects $\MAXAR_i$. Such a label $\ell_j$ always exists; it might be $\ell_i$ itself. 

\paragraph{Bounding the charges:} 
%\herman{Not sure what you mean with ``Scenario'' here. How about ``Bounding the charges''?}
Now consider a given event $e_j$. Let ${\cal C}_j$ be the set of events that charge to $e_j$. We will calculate an upper bound on $\sum_{e_i \in {\cal C}_j} \weightActR(\OPTAR_i)$, that is, the total charge to $e_j$, summed over all events in ${\cal C}_j$. \hermanchange{In fact, we will calculate an upper bound on
$\sum_{e_i \in {\cal C}_j} \weightEvent_i \cdot \size(\tau_i)$ that holds for any valid solution, and which is therefore also an upper bound on $\sum_{e_i \in {\cal C}_j} \weightActR(\OPTAR_i)$.} 

\begin{figure}
    \centering
    \includegraphics[page=3]{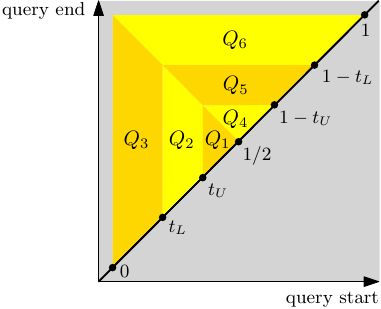}
    \caption{Illustration of \hermanchange{the different regions in the charging scheme.}}
    %case 1,2, and 3 of the case distinction for proving the upper bound of the approximation ratio. Queries $q_1$, $q_2$, and $q_3$ correspond to case 1, case 2, and case 3, respectively.}
\label{fig:approxratio-upperbound}
\end{figure}

\hermanchange{To derive this bound, we divide the triangle that contains the activity diagram into six regions $Q_1,...,Q_6$, calculate a bound on $\sum_{e_i \in {\cal C}_j} \weightEvent_i \cdot \size(\tau_i \cap C_h)$ for each $h \in \{1,...,6\}$, and add up the bounds. The six regions are determined as follows.} 
%\herman{Notation clash: $i$ is not defined. I guess we could write $\sum_{e_i \in {\cal C}_j} \weightActR(\OPTAR_i)$ but that has an unreadable double subscript. I would rather be able to say $\sum_{e \in {\cal C}_j} \OPTAR(e)$; then we need to define that notation somewhere}
% OLD 
%Now consider a given label $\ell_j$. Let ${\cal C}$ be the set of labels $\ell_i$ that charge to $\ell_j$. We will calculate an upper bound on the total charge to $\ell_j$, summed over all labels $\ell_i \in {\cal C}$. Note that for all labels $\ell_i \in {\cal C}$, the label $\ell_j$ is the first label selected by the greedy heuristic such that $\ell_j$ interferes with $\ell_i$: before $\ell_j$ is selected, each label $\ell_i \in {\cal C}$ could still get $\MAXAR_i$ as its activity region. However, as the greedy heuristic selects $\ell_j$ instead, we know we must have $\weightActR(\MAXAR_i) \leq \weightActR(\Greedy_j)$.
\hermanchange{let $t_L \leq t_U \leq 1/2$} be such that 
\begin{equation*}
    t_L (1-t_L) = \frac{1}{b} \cdot \size(\Greedy_j) \hspace{3em} \textrm{and} \hspace{3em}
    t_U (1-t_U) = b \cdot \size(\Greedy_j)
\end{equation*}
if such a $t_U$ exists \hermanchange{(that is, if $b \cdot \size(\Greedy_j) \leq 1/4$)}; otherwise $t_U = 1/2$. \hermanchange{Note:\[
\frac{t_U}{t_L} \leq 
\frac{b \cdot \size(\Greedy_j) / (1-t_U)}{\frac{1}{b}\cdot\size(\Greedy_j) / (1-t_L)} = b^2 \frac{1-t_L}{1-t_U} < b^2 \frac{1}{1/2} = 2b^2,
\]
and therefore:} $t_U < 2 b^2 t_L$. 
\hermanchange{Using $t_L$ and $t_U$, we can now define the six regions $Q_1,...,Q_6$ as illustrated in}%
%In the following, we distinguish different cases of time-window queries; see
~\autoref{fig:approxratio-upperbound}.
%\todo{Herman edited the figure to show all six regions; changed colour to improve contrast.}
\hermanchange{The idea of this subdivision in regions is that it distinguishes between three types of regions in the activity diagram. In $Q_1$ and $Q_4$ we find events whose maximum possible activity regions would have so much volume, that they would be selected before $e_j$ and therefore cannot charge to $e_j$. On the other extreme, charges from $Q_3$ and $Q_6$ are possible, but small, because these regions are too narrow to carry much volume. In between there are the regions $Q_2$ and $Q_5$, which cover charges to $e_j$ from events $e_i$ with maximum possible active regions similar to $e_i$. We will now analyze $\sum_{e_i \in {\cal C}_j} \weightEvent_i \cdot \size(\tau_i \cap C_h)$ for $h = 1, 2, 3$ in detail; the analysis for $h = 4, 5, 6$ is symmetric.}

\paragraph{Case $h=1$} 
%\herman{Ich habe $Q$ (für ein Query) durch $q$ ersetzt, damit man $Q$ for a set of queries benutzen kann}
First consider a time-window query $q = (t',t'')$ in the activity diagram with $t' + t'' \leq 1$ (therefore, $t' \leq 1/2$) and $t' > t_U$, \hermanchange{that is, a point $q \in Q_1$} in \autoref{fig:approxratio-upperbound}. Now, we look at an event $e_i$ that is active at $q$. For such an event $e_i$ it holds that $t_U<t_i <1-t_U$. Hence, we have $\size(\MAXAR_i) = t_i(1-t_i)$ and it holds that $t_i(1-t_i) > t_U (1-t_U)$. \hermanchange{Thus we find:}
\begin{align*}
&\size(\MAXAR_i) = t_i(1-t_i) > t_U (1-t_U) = b \cdot \size(\Greedy_j) \\
\Rightarrow
&\weightActR(\MAXAR_i) \geq (\weightEvent_j/b) \cdot \size(\MAXAR_i) > (\weightEvent_j/b) \cdot b \cdot \size(\Greedy_j) = \weightActR(\Greedy_j).
\end{align*}
Since we assumed that $e_i$ charges to $e_j$ which implies $\weightActR(\MAXAR_i) \leq \weightActR(\Greedy_j)$, such an event $e_i$ cannot be in ${\cal C}_j$. Otherwise, the greedy heuristic would have given $e_i$ its active region before $e_j$.
Thus, the total \eweight\ of labels in ${\cal C}_j$, intersected with the set of points $Q_1 = \{ (t',t'') \mid 0 \leq t' \leq t'' \leq 1 \mbox{ and } t'+ t'' \leq 1 \mbox{ and } t_U \leq t' \}$, is:\[
\sum_{e_i \in {\cal C}_j} \weightEvent_i \cdot \size(\tau_i \cap Q_1) = 0. 
\]

\paragraph{Case $h=2$} 
Consider a time-window query $q = (t',t'')$ with $t' + t'' \leq 1$ and $t_L \leq t' \leq t_U$, \hermanchange{that is, a point $q \in Q_2$} in \autoref{fig:approxratio-upperbound}. In any valid solution, the total number of events in ${\cal C}_j$ (whose labels all intersect $\ell_j$) that are active at $q$ is, by definition, at most~$a$. Furthermore, the \uweight\ \hermanchange{$\weightActR(\Greedy_i) / \size(\MAXAR_i)$} of any such event $e_i$ must be at most $\weightActR(\Greedy_j) / \size(\MAXAR_i) \leq \weightActR(\Greedy_j) / (t'(1-t'))$. 
%\todo[inline]{Q: die aussage verstehe ich nicht. A: Regarding the first part: if the weight would be more, then the maximum effective weight of $e_i$ would be more than $\weightActR(\Greedy_j) / \size(\MAXAR_i) \cdot \size(\MAXAR_i) = \weightActR(\Greedy_j)$, and this would still have been realizable when the greedy algorithm got to choose between $e_j$ and $e_i$, since $e_j$ is by definition the first label that gets in the way of $e_i$. So the greedy algorithm would have chosen $e_i$, not $e_j$ at that point. Regarding the second part: $\weightActR(\Greedy_j) / \size(\MAXAR_i) \leq \weightActR(\Greedy_j) / (t'(1-t'))$ holds because $\size(\MAXAR_i) = t_i(1-t_i) \geq t'(1-t')$ holds because $t' \leq t_i \leq 1/2$.}
Therefore, the total \uweight\ of the events active at $q$ is at most $a \cdot \weightActR(\Greedy_j) / (t'(1-t'))$. Thus, the total \eweight\ of the activity regions of labels in ${\cal C}_j$, intersected with the set of points $Q_2 = \{ (t',t'') \mid 0 \leq t' \leq t'' \leq 1 \mbox{ and } t'+ t'' \leq 1 \mbox{ and } t_L \leq t' \leq t_U \}$, is at most
\begin{eqnarray*}
\sum_{e_i \in {\cal C}_j} \weightEvent_i \cdot \size(\tau_i \cap Q_2)  
&= & a \cdot \weightActR(\Greedy_j) \cdot \int_{t_L}^{t_U} \left(\int_{t'}^{1-t'} \frac{1}{t'(1-t')}\,dt''\right)\,dt' \\
&= & a \cdot \weightActR(\Greedy_j) \cdot \int_{t_L}^{t_U} \frac{1-2 t'}{t'(1-t')}\,dt' \\
&= & a \cdot \weightActR(\Greedy_j) \cdot \int_{t_L}^{t_U} \left( \frac{1}{t'} - \frac{1}{1-t'} \right) dt' \\
&< & a \cdot \weightActR(\Greedy_j) \cdot \int_{t_L}^{t_U} \frac{1}{t'} dt' \\
&= & a \cdot \weightActR(\Greedy_j) \cdot \ln \frac{t_U}{t_L} \\
&< & a \cdot \weightActR(\Greedy_j) \cdot \ln (2b^2) \\
&= & a \cdot \weightActR(\Greedy_j) \cdot (\ln 2 + 2 \ln b). 
\end{eqnarray*}
%\todo[inline]{ist dies $\sum_{e_i \in \cal C} \weightActR(\OPTAR_i)$ oder $\sum_{e_i \in \cal C} \weightActR(\Greedy_i)$. Keines von beiden, denn es ist jetzt nur das Gewicht der active regions beschränkt auf dem Bereich $Q$:}

\paragraph{Case $h=3$} 
Finally, consider a time-window query $q = (t',t'')$ with $t' + t'' \leq 1$ and $t' < t_L$; \hermanchange{that is, a point $q \in Q_3$} in \autoref{fig:approxratio-upperbound}. We observe that they all lie in the trapezoid with vertices $(0,0),(t_L,t_L),(t_L,1-t_L),(0,1)$, whose size is $\hermanchange{t_L(1-t_L)=} \size(\Greedy_j)/b$, and the \uweight\ of any event of ${\cal C}_j$ active at $q$ is at most $b \cdot \weightEvent_j = b \cdot \weightActR(\Greedy_j) / \size(\Greedy_j)$. Thus, the total \eweight\ of the activity regions of the labels in ${\cal C}_j$, intersected with the set of points $Q_3 = \{ (t',t'') \mid 0 \leq t' \leq t'' \leq 1 \mbox{ and } t' + t'' \leq 1 \mbox{ and } t' < t_L \}$ is at most 
\begin{equation*}
 \sum_{e_i \in {\cal C}_j} \weightEvent_i \cdot \size(\tau_i \cap Q_3) = 
   a \cdot \frac{\size(\Greedy_j)}{b} \cdot \frac{b \cdot \weightActR(\Greedy_j) }{\size(\Greedy_j)} = a \cdot \weightActR(\Greedy_j).
\end{equation*}

\paragraph{Adding it up}
%For points $q = (t',t'')$ with $t' + t'' \geq 1$, the same analysis as for the cases 1, 2 and 3 goes through by symmetry. In particular, let the regions of the activity diagram considered in these cases be:\begin{eqnarray*}
%Q_4 & := & \{ (t',t'') \mid 0 \leq t' \leq t'' \leq 1 \mbox{ and } t'+ t'' \geq 1 \mbox{ and } t' < 1 - t_U \} \\
%Q_5 & := & \{ (t',t'') \mid 0 \leq t' \leq t'' \leq 1 \mbox{ and } t'+ t'' \geq 1 \mbox{ and } 1 - t_U \leq t' \leq 1 - t_L\\
%Q_6 & := & \{ (t',t'') \mid 0 \leq t' \leq t'' \leq 1 \mbox{ and } t' + t'' \geq 1 \mbox{ and } t' > 1 - t_L \}\\
%\end{eqnarray*}
Together, the sets of points $Q_h$ considered in cases 1, 2 and 3 and the symmetric cases cover the entire active diagram, that is:\[
\{ (t',t'') \mid 0 \leq t' \leq t'' \leq 1 \} = \bigcup_{h=1}^6 Q_h.
\]
Thus, in any valid solution, the total \eweight\ of the activity regions of the labels in ${\cal C}_j$ is at most 
\begin{eqnarray*}
\sum_{e_i \in {\cal C}_j} \weightActR(\tau_i) =
\sum_{e_i \in {\cal C}_j} \weightEvent_i \cdot \size(\tau_i) & = & 
\sum_{h=1}^6 \sum_{e_i \in {\cal C}_j} \weightEvent_i \cdot \size(\tau_i \cap Q_h) \\
& \leq &
2 \cdot a \cdot \weightActR(\Greedy_j) \cdot (\ln 2 + 2 \ln b) +
2 \cdot a \cdot \weightActR(\Greedy_j) \\
& = &
a \cdot \weightActR(\Greedy_j) \cdot (2 \ln 2 + 4 \ln b + 2). 
\end{eqnarray*}
This holds for any valid solution, so it also holds for the optimal solution $\OPTT(E)$ and we get:\[
\sum_{e_i \in {\cal C}_j} \weightActR(\OPTAR_i) 
\leq a \cdot \weightActR(\Greedy_j) \cdot (2 \ln 2 + 4 \ln b + 2).
\]
Note that each event $e_i$ charges to only one event $e_j$, and thus occurs in only one set ${\cal C}_j$. Thus we find:\[
\sum_{e_i \in E} \weightActR(\OPTAR_i) 
=
\sum_{e_j \in E} \sum_{e_i \in {\cal C}_j} \weightActR(\OPTAR_i) 
\leq 
a \cdot (2 \ln 2 + 4 \ln b + 2) \cdot \sum_{e_j \in E} \weightActR(\Greedy_j).
\]
%\textbf{Approximation Ratio:} 
This concludes the proof of approximation ratio $O(a \log b)$.
%\herman{Now we can refer to the lower bound and state that the bound is tight up to constant factors, that is, the approximation ratio is $\Theta(a \log b)$.} 
\annikachange{Thus, the approximation ratio is $\Theta(a \log b)$ and hence, we prove \autoref{thm:greedyapxratio}.}
Note that the approximation ratio can never be worse than $n$, as the first active region chosen by the greedy heuristic has at least as much \eweight\ as any active region in the optimal solution. Moreover, if all labels have equal \uweight\ ($b = 1$), the ratio $t_U/t_L$ in the above calculation becomes 1, that is, the \hermanchange{cases $h=2$ and $h=5$} disappear, and thus, the terms $(2 \ln 2 + 4 \ln b)$ disappear from the final bound.

\begin{corollary}
If all labels are unit squares of equal \uweight, the approximation ratio is at most 8. If all labels are unit disks of equal \uweight, the approximation ratio is at most 10.
\end{corollary}
\begin{proof}
The maximum number of mutually disjoint unit squares or disks that can intersect a given unit square or disk, respectively, is at most four or five, respectively.
\end{proof}

\bibliographystyle{abbrv-doi}

\bibliography{paper}
\end{document}